\documentstyle[12pt,a4]{article}

\def\be{\begin{equation}}
\def\ee{\end{equation}}  
\def\bea{\begin{eqnarray}}
\def\eea{\end{eqnarray}}
\def\vk{{\bf k}}
\def\vko{{\bf k}_{1}}
\def\vkt{{\bf k}_{2}}

\begin{document}
\begin{titlepage}
\begin{centering}

\vfill

{\bf Two-photon correlations as a sign of sharp transition
in quark-gluon plasma}.

\vspace{1cm}
I.V.Andreev\\
\vspace{0.5cm}
{\it
 Lebedev Physical Institute, 117924, Moscow, Russia}\\

\vspace{3cm}
\end{centering}
\begin{centerline}
{\bf Abstract}
\end{centerline}

\vspace{0.3cm}
The photon production arising due to time variation of the medium has been
considered. The Hamilton formalism for photons in time-variable medium
(plasma) has been developed with application to inclusive photon production.
The results have been used for calculation of the photon production in the
course of transition from quark-gluon phase to hadronic phase in relativistic
heavy ion collisions. The relative strength of the effect as well as
specific two-photon correlations have been evaluated. It has been demonstrated
that the opposite side two-photon correlations are indicative of the sharp
transition from the quark-gluon phase to hadrons.  
                                          
\vfill \vfill

\end{titlepage}

\section{Introduction}

The formation of the quark-gluon plasma (QGP) in relativistic heavy ion 
collisions with subsequent transition to hadrons is under discussion for 
many years. Numerous measurable signals from QGP phase (such as $J/\psi$ 
suppression) have been suggested. At the same time, the observation
of enhanced antibaryon production~\cite {BA} together with lattice
calculations~\cite{LAT} which predict low temperature of the transition 
are indicative of the fast QGP-hadron transition~\cite{CMM} without formation
of the mixed phase. In this paper we consider a new specific mechanism of photon
production which is effective in the case of sharp transition from QGP to
hadrons, see~\cite{A3}.

The phenomenon under consideration is the photon production in the course of
evolution of strongly interacting matter. Let us consider photons
in the medium at initial moment $t_0$ having momentum $\bf k$ and energy
$\omega_{in}$. Let the properties of the medium (its dielectric penetrability)
change within time interval $\delta\tau$ so that the final photon energy is
$\omega_f$. As a result of the energy change the production of extra photons
with momenta $\pm\bf k$ takes place these photons having specific
two-photon correlations. Moreover the photons with opposite momenta and
opposite helicities are produced even in the absence of the initial photons.
Analogous processes were considered for mesons~\cite{AW,AC,A,ACG} and applied
to pion production in high-energy heavy ion collisions~\cite{A1}. 
The conditions for strong effect are the following: first, the ratio of
the energies ${\omega_{in}/{\omega_f}}$ must not be too close to unity and
second, the transition should be fast enough.

The calculation of the effect requires consideration of the Hamilton
equations of motion. So in Section~2 we present in short the corresponding
formalism which is extended in Section~3 to the case of time-dependent
medium. A simple method of calculation of the photon correlations is 
presented in Section~4 with subsequent estimation in Section~5 of the
polarization operator and the evolution parameter which determines the
the strength of the transition effect. In the last Section~6 we calculate 
the transition effects in heavy ion collisions and present the results
of the work.

\section{Basic formulation}
We are interested in time evolution of photon creation and annihilation
operators $a^{\dag}_{i}(\vk,t)$ and $a_{i}(\vk,t)$ in the medium. In this
section the medium is taken in its rest frame. The properties of the medium 
(plasma) will be described by the transverse polarization operator 
$\Pi(\omega,\vk,T,\mu,m)$ which depends on energy $\omega$, momentum $\vk$,
temperature $T=1/\beta$, chemical potential $\mu$ and the mass $m$ of the
charged particles of the plasma. To get the evolution law of
$a^{\dag}_{i}(\vk,t)$ and $a_{i}(\vk,t)$ we will use the Hamilton formalism.

The Lagrangian of the electromagnetic field in the medium is taken in the
form
\be
L=\frac{1}{2}\int d^{3}x \left({\bf E}({\bf x},t){\hat\epsilon}{\bf E}({\bf x},t)
-{\bf H}^{2}({\bf x},t)\right)
\label{eq:1}   
\ee
where dielectric penetrability $\hat\epsilon$ acts as the factor
$\epsilon(\omega,\vk)$ in the momentum space (otherwise it acts as an operator).
The object of quantization is the real valued vector potential
${\bf A}({\bf x},t)$. In our case we can use the gauge conditions
\be
A_{0}({\bf x},t)=0, \qquad  div{\bf A}({\bf x},t)=0         
\label{eq:2}
\ee
so that
\be
{\bf E}({\bf x},t) =-{\bf \dot A}({\bf x},t),  \qquad
{\bf H}({\bf x},t)=rot{\bf A}({\bf x},t)
\label{eq:3}
\ee
The second of Eqs.(2) means that the vector potential is transverse one
$({\bf kA}=0)$ having two components $A_{i}$ (usual linear polarizations).
The transverse dielectric penetrability $\hat\epsilon$ is connected with
photon transverse polarization operator $\hat\Pi$ through equation
\be
\epsilon(\omega,k)=1-\Pi(\omega,k)/\omega^{2}
\label{eq:4}
\ee
where $\Pi(\omega,k)$ will be cosidered as the real valued and even
function of $\omega $ and $k$ (see Section 5).
Turning to momentum representation
\be
{\bf A}({\bf x},t) = \int\frac{d^{4}k}{(2\pi)^{2}}e^{-ik_{0}t+i{\bf kx}}
{\bf A}({\bf k},k_{0})
\label{eq:5}
\ee
and using (1-5) the action of the system can be written in the form
\be
S=\frac{1}{2}\int d^{4}k A_{i}(-\vk,-k_{0})\left[ k_{0}^{2}-{\vk}^{2}
-\Pi({\vk},k_{0})\right] A_{i}({\vk},k_{0})
\label{eq:6}
\ee
Variation of the action provides the equation
\be
\left[ k_{0}^{2}-{\vk}^{2}
-\Pi({\vk},k_{0})\right] A_{i}({\vk},k_{0}) =0, \qquad i=1,2
\label{eq:7}
\ee
and the energy $\omega$ of the photon in the medium is given by the usual 
dispersion equation
\be
\omega^{2}-{\vk}^{2}-\Pi(\omega,\vk)=0
\label{eq:8}
\ee
Let us note in advance that $\Pi(\omega,k)$ will be positive and slowly varying 
function of $k$ playing essentially the role of the effective photon mass
squared.

Introducing the time and momentum dependent field coordinates ${\bf q}(\vk,t)$
\be
{\bf A}({\bf x},t)=\int\frac{d^{3}k}{(2\pi)^{3/2}}e^{i{\bf kx}}{\bf q}(\vk,t),
\qquad {\bf q}(-\vk)={\bf q}^{\dag}(\vk)
\label{eq:9}
\ee
and coming one step back in (6) we get the Lagrangian in the form
\be
L=\frac{1}{2}\int d^{3}k\left[ \dot q_{i}(-\vk,t)\dot q_{i}(\vk,t)-{\vk}^{2}
q_{i}(-\vk,t)q_{i}(\vk,t) -q_{i}(-\vk,t)\hat \Pi(\vk) q_{i}(\vk,t)\right]
\label{eq:10}
\ee
where the polarization operator acts either to the left or to the right
giving equivalent results. In this form the polarization term is related 
to potential energy (unlike $\dot q$ terms) ensuring the physically sensible
form of the Hamiltonian in the case under consideration.
The Lagrange equations are given by variation of (10). Evidently they have
the oscillator form
\be
\ddot q_{i}(\vk)+\omega^{2}q_{i}(\vk)=0
\label{eq:11}
\ee
with $\omega$ determined from (8). That is
\be
\hat\Pi(\vk)q_{i}(\vk)=\Pi(\omega,\vk)q_{i}(\vk)
\label{eq:12}
\ee 
for plane wave solutions of the equations of motion.

Turning to Hamilton formalism for quantum fields, we introduce the canonically
conjugated momentum
\be
p_{i}(\vk,t)=\frac{\delta L}{\delta\dot q_{i}(\vk,t)}=\dot q_{i}(-\vk,t),
\qquad i=1,2
\label{eq:13}
\ee
and postulate canonical equal time commutation relations
\be
[q_{i}(\vko,t),p_{j}(\vkt,t)]=i\delta_{ij}\delta(\vko-\vkt)
\label{eq:14}
\ee
with all other commutators being zero. Let us note that the presence of the
opposite sign of $\vk$ in the {\sl rhs} of (13) is necessary also for
compatibility of (14) with commutation relations in coordinate space
\be
\left[A_{i}({\bf x}_{1},t),\dot A_{j}({\bf x}_{2},t)\right]=
i\delta_{ij}\delta({\bf x}_{1}-{\bf x}_{2})
\label{eq:15}
\ee
Below this opposite sign of the momentum $\vk$ will result in important
physical consequences describing production of photon pairs with opposite
directions of momenta of the photons.

The Hamiltonian is introduced in the usual way
\bea
H=\int d^{3}k p_{i}(\vk)\dot q_{i}(\vk) - L  \nonumber \\
=\frac{1}{2}\int d^{3}k \left( p_{i}(-\vk)p_{i}(\vk)
+ \vk^{2}q_{i}(-\vk)q_{i}(\vk) + q_{i}(-\vk)\hat\Pi(\vk)q_{i}(\vk)\right)
\label{eq:16}
\eea
and the Hamilton equations for Heisenberg operators
\bea
\dot q_{i}(\vk) = i\left[ H,q_{i}(\vk)\right] = p_{i}(-\vk), \nonumber \\
\dot p_{i}(-\vk) = i\left[ H,p_{i}(-\vk)\right] = - \omega^{2}q_{i}(\vk)
\label{eq:17}
\eea
connect $q_{i}(\vk)$ with $p_{i}(-\vk)$ being consistent with (11),(12).

Let us at last introduce the photon creation and annihilation operators
$a^{\dag}(\vk)$ and $a(\vk)$ through decomposition
\be
A_{i}({\bf x},t) = \int \frac{d^{3}k}{(2\pi)^{3/2}}\frac{1}{(2\omega)^{1/2}}
\left( a_{i}(\vk,t)e^{i{\bf kx}} +a_{i}^{\dag}(\vk,t)e^{-i{\bf kx}}\right)
\label{eq:18}
\ee
They are defined here for stationary medium in initial and final states
with constant $\omega$ and they are connected in these regions with 
canonical coordinates and momenta by equations
\bea
q_{i}(\vk) =\frac{1}{\sqrt{2\omega}}\left(a_{i}(\vk)+a_{i}^{\dag}(-\vk)\right) \nonumber \\
p_{i}(-\vk)=i\sqrt{\frac{\omega}{2}}\left(a_{i}^{\dag}(-\vk)-a_{i}(\vk)\right) 
\label{eq:19}
\eea
as it can be seen from comparison of representations (9) and (18) and their
time derivatives. The creation and annihilation operators have simple time
dependence
\be
a(t) \sim e^{-i\omega t}, \qquad a^{\dag}(t) \sim e^{i\omega t}
\label{eq:20}
\ee
corresponding to running waves (photons) and satisfy commutation relations
\be
\left[ a_{i}(\vko),a_{j}^{\dag}(\vkt)\right]=\delta_{ij}\delta(\vko-\vkt)
\label{eq:21}
\ee
which follow from canonical commutation relations (14).

\section{Photon evolution in time dependent \\medium}
Let the polarization operator be the time dependent function of its parameters.
In this case the Hamilton equations (17) remain valid having time dependent
energy $\omega(t)$ (as well as Lagrange equation (11) which is their
consequence). This is confirmed by the fact that their solution represents the
time dependent canonical transformation conserving commutator (14). Indeed,
the solution of (17) can be written in the form ({\sl cf}~\cite{BZP}):
\bea
q_{i}(\vk,t) = s_{1}(t)q_{i}(\vk,0) + s_{2}(t)p_{i}(-\vk,0) \nonumber \\
\dot q_{i}(\vk,t) = p_{i}(-\vk,t) = \dot s_{1}(t)q_{i}(\vk,0)
+ \dot s_{2}(t)p_{i}(-\vk,0)
\label{eq:22}
\eea
where $s_{1}(t)$ and $s_{2}(t)$ are two linearly independent real valued
solutions of the classical equation (11) with time dependent energy $\omega$
and with initial conditions
\be
s_{1}(0)=1,\quad \dot s_{1}(0)=0,\quad s_{2}(0)=0,\quad \dot s_{2}(0)=1
\label{eq:23}
\ee
One can see from (22) that the canonical commutator transforms in the following
way:
\be
\left[ q_{i}(\vk,t),p_{i}(\vk,t)\right] = W(s_{1},s_{2})
\left[ q_{i}(\vk,0),p_{i}(\vk,0)\right]
\label{eq:24}
\ee
where
\be
W(s_{1},s_{2}) = s_{1}(t)\dot s_{2}(t) - s_{2}(t)\dot s_{1}(t) 
\label{eq:25}
\ee
is the Wronskian determinant of (11) which does not depend on time for this
equation (due to absence of the term with the first order derivative in
the equation), this constant determinant being equal $1$ due to initial
conditions(23).

Let us consider the process of evolution of the photons from the initial
state with energy $\omega_{1}$ to asymptotic final state with energy
$\omega_{2}$. The final state annihilation and creation operators
are given by
\bea
a_{i}(\vk,t) = \frac{1}{\sqrt{2\omega_{2}}}\left( \omega_{2}q_{i}(\vk,t)+
ip_{i}(-\vk,t)\right) \nonumber \\
a_{i}^{\dag}(\vk,t) = \frac{1}{\sqrt{2\omega_{2}}}\left( \omega_{2}q_{i}(-\vk,t)
-ip_{i}(\vk,t)\right)
\label{eq:26}
\eea
as it follows from (19). We substitute solutions (22) for $q_{i}(\vk,t),
p_{i}(\vk,t)$ and introduce two lineary independent complex valued classical
solutions $\xi(t), \xi^{*}(t)$ instead of $s_{1}(t), s_{2}(t)$
\be
\xi(t) = s_{1}(t)+i\omega_{1}s_{2}(t),\quad
\xi^{*}(t) = s_{1}(t)-i\omega_{1}s_{2}(t) 
\label{eq:27}
\ee
with initial conditions
\be
\xi(0) = \xi^{*}(0) = 1,\quad \dot \xi(0) = i\omega_{1},\quad 
\dot\xi^{*}(t) = -i\omega_{1}
\label{eq:28}
\ee
Then, using (19) for the initial state, we get the Bogoliubov
transformation~\cite{B}
connecting the creation and annihilation operators in the initial and final
states (let us remind that these operators were defined only for asymptotic
states having constant energy $\omega$):
\bea
a_{i}(\vk,t) = u(\vk,t)a_{i}(\vk,0) + v(\vk,t)a_{i}^{\dag}(-\vk,0) \nonumber \\
a_{i}^{\dag}(\vk,t) = v^{*}(\vk,t)a_{i}(-\vk,0) + u^{*}(\vk,t)a_{i}^{\dag}(\vk,0)
\label{eq:29}
\eea
with
\bea
u(\vk,t) = \frac{1}{2}\sqrt\frac{\omega_{2}}{\omega_{1}}\left[\xi^{*}(t)
      +\frac{i}{\omega_{2}}\dot \xi^{*}(t)\right] \nonumber \\
v(\vk,t) = \frac{1}{2}\sqrt\frac{\omega_{2}}{\omega_{1}}\left[\xi(t)
      +\frac{i}{\omega_{2}}\dot \xi(t)\right]
\label{eq:30}
\eea
It follows from (30) that
\be
u^{*}(t)u(t) - v^{*}(t)v(t) = \frac{i}{2\omega_{1}}W(\xi, \xi^{*})
\label{eq:31}
\ee
where
\be
W(\xi, \xi^{*}) = \xi(t)\dot \xi^{*}(t) - \dot \xi(t)\xi^{*}(t)
\label{eq:32}
\ee
is again the time independent Wronskian determinant. So
\be
u^{*}u -v^{*}v = 1
\label{eq:33}
\ee
due to initial conditions (28). In turn, as it follows from (29), the
condition (33) ensures conservation of the commutator
\be
\left[a_{i}(\vk,t), a^{\dag}_{i}(\vk,t)\right]
= \left[a_{i}(\vk,0), a^{\dag}_{i}(\vk,0)\right] 
\label{eq:34}
\ee
(the last commutator is equal to $V/(2\pi)^{3}$ for the system having volume 
$V$). In view of condition (33) the Bogoliubov coefficients $u,v$ can be
represented in the form
\be
u = \cosh r(\vk) e^{i\alpha_{1}},\quad v = \sinh r(\vk) e^{i\alpha_{2}}
\label{eq:35}
\ee
where $r(\vk)$ is the main parameter which determines the photon production
and the phases $\alpha_{1}, \alpha_{2}$ do not play important role and they 
will not be considered below.

To find the coefficients $u,v$ (for fixed momentum $\vk$) one must turn
to classical equations for oscillator having variable frequency (energy).
We look for solution of the equation 
\be
\ddot \xi + \omega^{2}(t)\xi = 0
\label{eq:36}
\ee
In the initial state, where the energy $\omega_{1}$ is constant, we take
a single wave
\be
\xi(t) = e^{i\omega_{1}t}, \quad t<t_{in}=0
\label{eq:37}
\ee
At large enough time, $t>t_{f}$, when the energy $\omega_{2}$  becoms constant
again, the general solution has the form
\be
\xi(t) = C_{1}e^{i\omega_{2}t} + C_{2}e^{-i\omega_{2}t}, \quad t>t_{f}
\label{eq:38}
\ee
Substituting (37),(38) into (30) we get corresponding Bogoliubov coefficients
\be
u = \sqrt\frac{\omega_{2}}{\omega_{1}}C^{*}_{1}e^{-i\omega_{2}t},\quad
v = \sqrt\frac{\omega_{2}}{\omega_{1}}C_{2}e^{-i\omega_{2}t},\quad t>t_{f}
\label{eq:39}
\ee
where constants $C_{1}, C_{2}$ satisfy relationship
\be
\mid C_{1}\mid^{2} - \mid C_{2}\mid^{2} = \frac{\omega_{1}}{\omega_{2}}
\label{eq:40}
\ee
due to condition (33).

To find the final expressions for coefficients $u,v$ one must know
the full solution of (36) connecting the asymptotics (37) and (38).
Here one can use an analogy between the above problem and the problem
of the wave propagation through (above) the one-dimensional potential 
barrier. In the last case (38) (after substitution of $x$ for $t$) represents
incoming and reflected waves and (37) represents outgoing wave. The
reflection coefficient
\be
\mid{C_{2}}/{C_{1}}\mid^{2} = \mid{v}/{u}\mid^{2} = \tanh^{2}r
\label{eq:41}
\ee
gives the desired ratio of the Bogoliubov coefficients (up to phases).
Therefore one can use known quantum-mechanical results. So, shifting
the initial time $t_{in}$ to large negative value, we take the reference
model of the energy variation:
\be
\omega^{2}(t) = \frac{\omega_{2}^{2}+\omega_{1}^{2}}{2} + 
\frac{\omega_{2}^{2}-\omega_{1}^{2}}{2}\tanh\left(\frac{2t}{\delta\tau}\right)
\label{eq:42}
\ee
The problem with such form of the potential barrier can be found in
textbooks~\cite{LL}. It contains the important parameter $\delta\tau$
giving characteristic time of the energy variation. The evolution parameter
$r$ is given now by
\be
r = \tanh^{-1}|v/u| = \frac{1}{2}\ln\left(\frac{\tanh(\pi\omega_{2}\delta
\tau/4)}{\tanh(\pi\omega_{1}\delta\tau/4)}\right)
\label{eq:43}
\ee
For sharp transition $(\omega_{i}\delta\tau \ll 1)$ we get from (43):
\be
r = \frac{1}{2}\ln\left(\frac{\omega_{2}}{\omega_{1}}\right), \quad
\left|\frac{u}{v}\right|^{2} = \left(\frac{\omega_{2}-\omega_{1}}
{\omega_{2}+\omega_{1}}\right)^{2}, \quad \omega_{i}\delta\tau\ll1  
\label{eq:44}
\ee
This is the case of the most intensive pair production. For large transition
time $\delta\tau$ the process becoms adiabatic one and the evolution
parameter falls down exponentially:
\be
r = \left|e^{-\pi\omega_{2}\delta\tau/2} - e^{-\pi\omega_{1}\delta\tau/2}
\right|, \quad \omega\delta\tau \gg 1
\label{eq:45}
\ee

Let us note that the above results (44) for sharp transition can be found
immediately from Hamilton equations (17). Suggesting that the canonical 
coordinate $q$ and momentum $p$ are finite, we see from these equations
that $\dot q$ and $\dot p$ are finite as well and so $q$ and $p$ are
continuous functions of time at the transition point:
\be
q(-\delta t) = q(\delta t),\quad p(-\delta t) = p(\delta t),\quad 
\delta t \to 0
\label{eq:46}
\ee
coinsiding for both sides of the sharp border between media with different
photon energies $\omega_1$ and $\omega_2$. Being expressed according (19)
through creation and annihilation operators $a^{\dag}, a$ at both sides
of the border,the equations (46) lead to Bogoliubov transformation (29)
between pairs of operators $a^{\dag}, a$ at different sides of the border
with coefficients
\be
|u| = \frac{\omega_{2}+\omega_{1}}{2\sqrt{\omega_{1}\omega_{2}}},\quad
|v| = \left|\frac{\omega_{2}-\omega_{1}}{2\sqrt{\omega_{1}\omega_{2}}}\right|
\label{eq:47}
\ee
The last equations correspond to (44).

\section{Inclusive photon production in a simple \\ model}
Below we will be interested in inclusive photon production in heahy ion
collisions. We confine ourselves to symmetric case when the photons having
opposite momenta ${\pm}{\vk}$ are produced in an equivalent way
(central collisions of identical nuclei). For simplicity the Bogoliubov
coefficients $u(\vk),v(\vk)$ will be taken to be real valued and $k=|\vk|$ 
dependent.
To get feeling of the main features of the photon production and their
correlations (and for further references and comparison) we formulate in this
section a simple model -- fast simultaneous transition of large homogeneous
system at rest (the movement of the system will be considered in Section 6).

For physical interpretation of the evolution effect it is helpful to 
introduce the complex valued vectors of the photon circular polarization:
\be
{\bf e}_{\pm}=({\bf e}_{1} {\pm} i{\bf e}_{2})/\sqrt2, \quad {\vk}{\bf e}_{\pm}=0
\label{eq:48}
\ee
and corresponding components of operators ${\bf a}(\vk),{\bf a^{\dag}}(\vk) $
\be
a_{\pm} = (a_{1} \pm ia_{2})/\sqrt2,           \quad
a_{\pm}^{\dag} = (a_{1}^{\dag}\mp ia_{2}^{\dag})/\sqrt2  
\label{eq:49}
\ee
so that
\be
{\bf a}(\vk)={\bf e}_{+}a_{-}(\vk)+{\bf e}_{-}a_{+}(\vk), \quad
a_{\pm}={\bf e}_{\pm}{\bf a}
\label{eq:50}
\ee
The components $a_{\pm},a_{\pm}^{\dag}$ satisfy the standard commutation relations
of the form (21) with $i={\pm}$ and represent the creation and annihilation
operators of the photons with definite spin projection ${\pm}1$ on the direction
of the photon momentum (the helicity).

In what follows we will denote the creation and annihilation operators
in the final state as $b^{\dag}, b$ leaving notations $a^{\dag},a$ for
the operators in the initial state. In terms of the components $b_{\pm}$
 the helicity operator is
\be
S_{b3}=i(b^{\dag}_{1}b_{2}-b^{\dag}_{2}b_{1})=
b^{\dag}_{+}b_{+}-b^{\dag}_{-}b_{-}=N_{b+}-N_{b-}
\label{eq:51}
\ee
for every momentum ${\vk}$, the photon number operator is
\be
 N_{b}=b^{\dag}_{+}b_{+}+b^{\dag}_{-}b_{-}=N_{b+}+N_{b-}
\label{eq:52}
\ee
and the Bogoliubov transformation takes the form:
\be
b_{\pm}(\vk)=ua_{\pm}(\vk)+va^{\dag}_{\pm}(-\vk), \quad
b^{\dag}_{\pm}(\vk)=u^{*}a^{\dag}_{\pm}(\vk)+v^{*}a_{\pm}(-\vk)
\label{eq:53}
\ee
where indexes $\pm$ are helicities related to corresponding momenta
$\vk$ or $-\vk$.
   
The resulting average number of the photons and their correlations depends
on the initial state. We suggest that the initial state is the statistical
(gaussian) mixture of the coherent  states~\cite{{GS},{AW}} of the photons of
each polarization.Then using (53) we get the photon momentum distributions
\be
\left(\frac{dN}{d^{3}k}\right)_{\pm}=\langle N_{b\pm}({\vk})\rangle=
\left(1+|v|^{2}\right)\langle N_{a\pm}({\vk})\rangle
+|v|^{2}\langle N_{a\pm}(-{\vk})\rangle+|v|^{2}\frac{V}{(2\pi)^{3}}
\label{eq:54}
\ee
                 
As one can see from (54) the photons are created in pairs having opposite
directions of their momenta and opposite spin directions (the same helicities).
If both kinds of polarizations are equally
represented in the initial state for every momentum ${\vk}$, then the same
property is valid for the final state,
\be
\langle S_{b3}({\vk})\rangle = 0
\label{eq:55}
\ee
and the photons produced with opposite momenta have equal helicities.
The intensity of the transition production is given by the Bogoliubov
coefficient $|v|^{2}$ and the last
term in {\sl rhs} of (54) represents the result of the ground state
rearrangement ( the initial ground state is not the ground state for
$b$-operators which operate in another medium in the final state).
We suggest that ${\vk}\to{-\vk}$ symmetry takes place. Then
the resulting photon momentum distribution can be written in the form:
\be
\frac{dN}{d^{3}k} =\langle b^{\dag}_{\lambda}({\vk})b_{\lambda}({\vk})\rangle
= \frac{2V}{(2\pi)^{3}}\left[ n({\vk}) +
\left(2n({\vk})+1\right)\sinh^{2}r \right]
\label{eq:56}
\ee
(sum over helicities ${\lambda}$)
where $n({\vk})$ is the average level occupation number of the single mode
in the initial state and we used the parametrization (35).
The photon production in the course of the transition is given by the
second term in {\sl rhs} of (56). It is weak for small evolution parameter
$r(k)$ being of the order $r^{2}$.

The evolution effect is better seen in photon correlations. Two-photon 
inclusive cross-section is given by
\bea
\frac{1}{\sigma}\frac{d^{2}\sigma}{d^{3}k_{1}d^{3}k_{2}} =
\langle b^{\dag}_{\lambda}({\vko})b^{\dag}_{\mu}({\vkt})b_{\lambda}({\vko})
b_{\mu}({\vkt})\rangle
=\langle b^{\dag}_{\lambda}({\vko})b_{\lambda}({\vko})\rangle
 \langle b^{\dag}_{\mu}({\vkt})b_{\mu}({\vkt})\rangle  \nonumber \\
+\langle b^{\dag}_{\lambda}({\vko})b_{\mu}({\vkt})\rangle
 \langle b^{\dag}_{\mu}({\vkt})b_{\lambda}({\vko})\rangle
+\langle b^{\dag}_{\lambda}({\vko})b^{\dag}_{\mu}({\vkt})\rangle
 \langle b_{\lambda}({\vko})b_{\mu}({\vkt})\rangle
\label{eq:57}
\eea
The first term in {\sl rhs} of (57) is the product of single-photon
distributions, the second term gives the Hanbury Brown-Twiss effect
(HBT, called also Bose-Einstein correlations) and the third term is essential
if the time evolution effect takes place giving opposite side photon
correlations.

Production of the photons with opposite directions of their momenta was already
noted after (54). However it was implicitely suggested there that the volume
of the system is arbitrary large, $V\to\infty$. In real situation we deal
with large but finite size of the colliding nuclei. The finite size of the
photon source will smooth out the effect. The same is valid for HBT effect
responsible for the same side correlations. The special technique was elaborated
in a number of papers to describe the source size effect for HBT correlations
including the use of Wigner phase space density~\cite{P} and the method of
equivalent classical currents~\cite{APW} which can be used here. Instead
we directly modify the creation and annihilation operators~\cite{AW} in such
a way that they are nonzero only inside some region in coordinate space
which is described by the function $f({\bf x})$:
\be
\tilde b_{\lambda}({\bf x}) = b_{\lambda}({\bf x})f({\bf x}), \quad
\tilde b^{\dag}_{\lambda}({\bf x}) = b^{\dag}_{\lambda}({\bf x})f({\bf x})
\label{eq:58}
\ee
Then the modified (smoothed out) operators in momentum space are:
\be
\tilde b^{\dag}_{\lambda}({\vko})=\int d^{3}kf({\vko-\vk})
 b^{\dag}_{\lambda}({\vk}),\quad
\tilde b_{\lambda}({\vkt})=\int d^{3}kf({\vk-\vkt})
 b_{\lambda}({\vk})
\label{eq:59}
\ee                 
where the Fourier transform $f({\bf k_{i}-\vk})$ is sharply peaked around
the point ${\vk}={\bf k}_{i}$ (smoothed $\delta$-function).

The introduced operators satisfy modified commutation relations
\bea
\left[\tilde b_{\lambda}({\vko}),\tilde b^{\dag}_{\nu}({\vkt})\right]_{-}=
\delta_{\lambda\nu}\int d^{3}kf({\vko-\vk})f({\vk-\vkt}) \nonumber \\
=\delta_{\lambda\nu}\int\frac{d^{3}x}{(2\pi)^{3}}f^{2}({\bf x})
e^{i({\vko-\vkt}){\bf x}} =\delta_{\lambda\nu}F({\vko-\vkt})
\label{eq:60}
\eea
Here $F(\vko-\vkt)$ is the form-factor of the photon source (Fourier transform
of the source density) and it also represents the smoothed $\delta$-function
which has the width of the order of the inverse size of the source.
In particular
\be
F(0) = V/(2\pi)^{3},
\label{eq:61}
\ee
where $V$ is the effective volume of the source at the stage of the transition
in the medium. Let us note that above we in fact already used (61) substituting
it for $\delta^{3}(0)$ in the contribution of the ground state rearrangement
to the evolution effect in (54).This factor appears also when one wants to use
the level population function $n({\vk})$ (see (56)), having its origin in the
correspondence of discrete and continous Fourier decompositions.
Using operators modified in three-dimensional coordinate space we suggest
that the volume $V$ changes inessentially in the course of the time transition.

Now one can apply the Bogoliubov transformation (53) to estimate the correlators
of the modified operators: 
\bea
\langle \tilde b^{\dag}_{\pm}({\vko})\tilde b_{\pm}({\vkt})\rangle =
\int d^{3}k\left[u^{2}(k)n(k)+v^{2}(k)(n(k)+1)\right]f({\vko-\vk})f({\vk-\vkt})
\nonumber \\
=\frac{V}{(2\pi)^{3}}\left[ n({\vk})+\left(2n({\vk})+1\right)\sinh^{2}r \right]
G(\vko-\vkt)
\label{eq:62}
\eea
and in a similar way
\bea
\langle \tilde b_{\pm}({\vko})\tilde b_{\pm}({\vkt})\rangle=
\langle \tilde b^{\dag}_{\pm}({\vko})\tilde b^{\dag}_{\pm}({\vkt})\rangle
\nonumber \\
=\frac{V}{(2\pi)^{3}}(2n(k)+1)\sinh r(k)\cosh r(k)G(\vko+\vkt)
\label{eq:63}
\eea
with other correlators vanishing in the course of statistical averaging
in the case of sharply peaked form-factor.
In the above equations $G(\vko\pm\vkt)$ is the normalized form-factor
$(G(0)=1)$ and we took into account that the functions $f({\vko-\vk})$ and
$f({\vk-\vkt})$ as well as form-factor $G(\vko\pm\vkt)$ are sharply peaked
functions of their arguments (at zero momentum) having characteristic
scale of the order of inverse size of the source, this scale being much less
than the characteristic scales of the momentum distribution $n(k)$ and
the evolution parameter $r(k)$. So the last two functions can be evaluated
at any of momenta $\vko,\vkt\approx\pm\vk $ (we suggest that the process
is $\vk\to\-\vk$ symmetric). The equations (62-63) together with (57)
show that the correlations arising due to photon identity (HBT effect)
are the same side momentum correlations of the photons having the same
helicities whereas the photons arising due to transition effect have
the opposite side momentum correlations and approximately opposite spin
directions (the same helicities again).

Returning to two-photon correlations given by (57) (sum over helicities)
we get the relative correlation function which is measured in experiment:
\be
C({\vko,\vkt})=1+\frac{1}{2}G^{2}({\vko-\vkt})+\frac{1}{2}R^{2}_{0}(k)
G^{2}({\vko+\vkt})
\label{eq:64}
\ee
with
\be
R_{0}(k)=\frac{(2n(k)+1)\sinh r(k)\cosh r(k)}
              {n(k)+(2n(k)+1)\sinh^{2}r(k)}
\label{eq:65}
\ee
according to (62-63). As can be seen from (64-65) the transition effect
depends strongly on the evolution parameter r(k),see(43-45). Below, after
necessary modifications, we apply the above considerations to photon
transition radiation in heavy ion collisions.

\section{Photons in plasma and the evolution \\ parameter}
To find the evolution parameter $r(k)$ one must know the photon energy in
plasma.The spectrum of photons in plasma is given by dispersion equation
\be
\omega^{2}_{k}=k^{2}+\Pi(\omega_{k},k,T,\mu,m)
\label{eq:66}
\ee
Here $\Pi$ is the polarization operator for transverse photons dependent on
temperature $T=\beta^{-1}$, chemical potential $\mu$ and the mass $m$ of charged 
particles. Below we use an approximate form of $\Pi$ extracted from original
expression~\cite{F}:
\be
\Pi(\omega,k)=\omega^{2}_{a}\left[ 1-\frac{\omega^{2}-k^{2}}{k^{2}}\left( \ln \left(
 \frac{\omega+vk}{\omega-vk} \right)-1 \right) \right]
\label{eq:67}
\ee
with
\be
\omega^{2}_{a}=\frac{4g\alpha }{\pi\beta^{2}}\int_{m\beta}^{\infty}dx
\left(x^{2}-{m^{2}\beta^{2}}\right)^{1/2}n_{F}(x,\mu\beta)
\label{eq:68}
\ee
where $\alpha=1/137$, $v^2$ is the averaged velocity squared of the charged
particles in the plasma, factor $g$ takes into account the number of the
particle kinds and their electric charges ($g=5/3$ for $u,d$ quarks) and
$n_{F}$ is the occupation number of the charged particles:
\be
n_{F}(\beta\omega)=\left( e^{\beta\omega-\beta\mu}+1\right)^{-1}+
                   \left( e^{\beta\omega+\beta\mu}+1\right)^{-1}
\label{eq:69}
\ee
(Fermi distribution).The polarization operator for scalar charged particles
is approximately a half of that for fermions with substitution of Bose
distribution for Fermi distribution. The imaginary part of the polarization
is small in the case under consideration (no Landau damping) and it was
neglected in the calculations.

For small masses of charged particles,$\beta m\ll1$, the asymptotics of the
polarization operator takes the known simple form
\be
\omega^{2}_{a}=\frac{2\pi g\alpha}{3}\left( T^{2}+\frac{3}{\pi^{2}}\mu^{2}
\right)
\label{eq:70}
\ee
This form can be used in quark-gluon phase. For large masses of fermions
(constituent quarks and nucleons) when $\beta m>1 (m>\mu) $ it is convenient
to use the expansion 
\be
\omega^{2}_{a(F)}=\frac{8g\alpha m}{\pi\beta}\sum^{\infty}_{n=1}
\frac{(-1)^{n-1}}{n}\cosh(n\beta\mu)K_{1}(n\beta m)
\label{eq:71}
\ee
where $K_{1}(x)$ is the exponentially decreasing modified Bessel function
and the main contribution comes from a few first terms of the series.
Analogous expansion was also used for bosons (pions) having no chemical
potential:
\be
\omega^{2}_{a(B)}=\frac{4\alpha m}{\pi\beta}\sum^{\infty}_{n=1}
\frac{1}{n}K_{1}(n\beta m)
\label{eq:72} 
\ee

We considered the polarization operator and photon spectrum for three possible
kinds of plasma: quark-gluon plasma (QGP) with $u,d$ light quarks, constituent
quark($m=350 MeV$)-pion plasma and hadronic (pions and nucleons) plasma.
Chemical potential (baryonic one) was taken to be equal to $100 MeV$ per quark
corresponding to typical value for CERN-SPS energies (say $160 GeV$ per
nucleon in $Pb-Pb$ collisions). The temperature $T_{c}$ of the transition
was taken to be equal to rather small value $140 MeV $ characteristic for
final hadrons~\cite{R} ($T_{c}=200 MeV $ was also tested giving close results). 
Under above conditions the asymptotic values $\omega_{a}$ in (70-72) are
equal to 24, 11.7, 11.2 and 2.5 in $MeV$ units for light quarks, constituent
quarks, pions and nucleons correspondingly.

Let us note that the approximation (67) where we use the averaged velocity
squared $v^{2}=\langle v^{2} \rangle$, suggests that higher moments of the
velocity distribution do not differ significantly from corresponding powers
of $v^{2}$. So we calculated the ratio $\langle v^{4} \rangle/\langle v^{2}
\rangle^{2}$ and verified that it differs from unity inessentially being 1
for QGP, 1.11 for constituent quarks (valons), 1.06 for pions and reaching 
1.25 for nucleons which last contribution is small by itself. Corresponding
values of $v^{2}$ used in our estimations are 1, 0.545, 0.731 and 0.300.

Evidently the polarization operator in (66) plays the role of (momentum
dependent) photon mass squared $m^{2}_{\gamma}$. The termal mass squared
at zero photon momentum $k$ is equal to
\be
\omega^{2}_{0}= \omega^{2}_{a}\left( 1 - \frac{v^{2}}{3} \right)
\label{eq:73}
\ee
and it approaches $\omega^{2}_{a} $ at large momenta.
The slope of the dispersion curve at the origin is:
\be
\frac{d\omega^{2}_{k}}{dk^{2}}\left|_{k=0}
=\left(1-\frac{1}{5}v^{2}\right)\right/\left(1-\frac{1}{3}v^{2}\right)=1+c
\label{eq:74}
\ee
varying from 1 at $v=0$ to 1.2 at $v=1$. In rough approximation the polarization
operator taken at the dispersion curve can be represented by the simple 
expression
\be
\Pi(\omega_{k},k)=m^{2}_{\gamma}(k)=\omega^{2}_{0}+\frac{c\omega^{2}_{0}k^{2}}
{\omega^{2}_{0}+dk^{2}}, \quad
d=1-\frac{2}{5}v^{2}
\label{eq:75}
\ee
reproducing position of the point $\omega_{0}$, asymtotic value $\omega_{a}$
and the slope (74). In the presence of two different sorts of charged particles
(say pions and nucleons) the expression (75) should be modified in an 
appropriate way to take into account the presence of two contributions and
to ensure the new correct slope $d\omega^{2}/dk^{2}$.
The simple approximation (75) appears to be rather close to polarization
operator (67) taken at the dispersion curve (66) (these two curves have a common
point also at some finite $k$) and it will be used below for estimation of the
evolution parameter $r(k)$. Zero momentum photon masses $m_{\gamma}(0)$
was found to be 19.5, 14.4 and 10.0 in $MeV$ units for QGP, valon-pion and
nucleon-pion plasma correspondingly.

The evolution parameter $r(k)$, which gives the strength of the transition
radiation, was determined in (43-45) for the reference model (42). It depends
strongly on transition time $\delta\tau$. We suggest that this time interval
is not large being of the order of $1 fm/c $. Then for small momenta $k$
($\omega(k)\delta\tau\ll 1$) the parameter $r(k)$ is universal and it can be
well approximated by the expression which follows from (44): 
\be
r(k)=\frac{m^{2}_{\gamma1}(k)-m^{2}_{\gamma2}(k)}
{4(\langle m^{2}_{\gamma}(k)\rangle+k^{2})}
=\frac{\delta m^{2}_{\gamma}(k)}{4\langle\omega^{2}(k)\rangle} ,
\qquad \omega(k)\delta\tau\ll1
\label{eq:76}
\ee
where $m^{2}_{\gamma i}$ are photon termal masses squared at both sides of
the transition and $\langle m^{2}_{\gamma}\rangle$ is their average. 
Zero momentum evolution parameter $r(0)$ is equal to 0.324, 0.154 and 0.178
for QGP-hadron, QGP-valon and valon-hadron transitions correspondingly.
Higher momentum behaviour of $r(k)$ is model dependent. Below it will be
taken in the form
\be
r(k)=\frac{\delta m^{2}_{\gamma}}{4k^2}\exp \left(-\frac{\pi}{2}k\delta\tau\right)
\label{eq:77}
\ee
The Eq.(77) is a simple version of the asymptotical form (45) and it can be
well sewed together with (76) giving a monotonically decreasing function of the
momentum. Let us note that on general grounds one expects that $r(k)$ falls
down exponentially at large $k\delta\tau$ if the time dependence of the energy
$\omega(k,t)$ in the course of the transition has no singularities
(non-analiticity) at real time axis.
 Below Eqs.(76-77) will be used for estimation of the transition effect
in heavy ion collisions. Only QGP-hadron transition will be considered.
In view of fact that the evolution parameter $r(k)$ appears to be small number  
at all momenta $k$, all expressios will be estimated in the lowest order
in $r(k)$.

\section{Transition effect in heavy ion collisions}
Let us now apply the above consideratins to photon production in heavy ion
collisions. Let us suggest that the quark-gluon plasma is formed at the
initial stage of the ion collision. Let the plasma undergoes expansion and
cooling. The expansion is taken to be longitudinal and boost invariant~\cite{BJ}.
Recent lattice calculations~\cite{LAT} show rather low critical temperature
of the deconfinement and chiral phase transition, $T_c\approx 150 MeV$
as well as sharp drop of the pressure up to very small value
when the temperature approachs $T_c$
thus provocing instability in the presence of overcooling. So we do not
expect long-living mixed phase and consider fast transition from quark to
hadron matter with characteristic transition proper time duration $\delta\tau$
of the order of $1 fm/c$.

To calculate the transition effect one must shift to rest frame of each moving
element of the system and integrate over proper times $\tau$ and space-time
rapidities $\eta$ of the elements of the system. Then the invariant
single-photon distribution in central rapidity region $y=0$ reads:

\begin{eqnarray}
\frac{dN}{d^{2}k_{T}dy}\Bigl|_{y=0}\Bigr.=I_{QGP}+I^{(1)}_{tr}\nonumber\\
=\int\tau d\tau\int d\eta\int d^{2}x_{T}(p_{0}\frac{dR_{\gamma}}{d^{3}p})
+\int d\eta\int d^{2}x_{T}\frac{2p\tau_{c}}{(2\pi)^{3}}r^{2}(p)
\label{eq:78}
\end{eqnarray}                       
with $p=k_{T}\cosh\eta$.

The first term in {\sl rhs} of (78) describes photon production from hot 
quark-gluon plasma. Here $R_{\gamma}$ is the QGP production rate per unit
four-volume in the rest frame of the matter~\cite{KLS}:

\begin{equation}
p_{0}\frac{dR_{\gamma}}{d^{3}p}=\frac{5\alpha\alpha_{s}}{18\pi^{2}}T^{2}
\exp(-p/T)\ln(1+\frac{\kappa p}{T})
\label{eq:79}
\end{equation}                            
with $\alpha=1/137$, $\alpha_{s}=0.4$, $\kappa=0,58$. It can be used  also for
hadron gas as its uncertainity is larger than the difference between the
first-order QGP and hadron gas production rates~\cite{NKL}. Contribution from 
hadronic resonances are not considered here. The second term in {\sl rhs} of
(78)
describes photon production due to transition from QGP to hadrons in the
vicinity of proper time $\tau_{c}$,{\sl cf } (56). The time duration of the
transition is taken to be small in this term in comparison with total time
duration of photon production process. The evolution parameter $r(p)$ is
given here by (76-77).

As the last step one must specify the temperature evolution in (78).
We suggest that the temperature depends on proper time of the volume element
with power-like dependence:
\begin{equation}
(T/T_{0})=(\tau/\tau_{0})^{-1/b}
\label{eq:80}
\end{equation}
where $\tau_{0}$ and $T_{0}$ are initial proper time and initial temperature.
For final estimation we use $b=3$ typical for hydrodynamical picture and
choose low transition temperature $T_{c}=140 MeV$. After transition the photons
live some time in hadronic medium and we suggest termal momentum distribution
of the hadrons (modified by the expansion of the system). We neglect termal
photon production below $T_{c}$ and do not introduce a special freese-out
temperature. An alternative way is to introduce the final temperature  
$T_{f}<T_{c}$. Not pretending to high accuracy we shall not distinguish
between $T_{c}$ and $T_{f}$ below.

Using now the (79-80) the photon production rate in QGP phase in (78)
can be integrated over space-time rapidities $\eta$:
\begin{equation}
I_{QGP}=\frac{5\alpha\alpha_{s}}{9\pi^{2}}(2\pi)^{1/2}b\frac{(k_{T}\tau_{c})^2}
{(k_{T}\beta_{c})^{2b-2}}\int d^{2}x_{T}J_{QGP}
\label{eq:81}
\end{equation}
with
\begin{equation}
J_{QGP}\cong\int_{k_{T}\beta_{0}}^{k_{T}\beta_{c}}dx\frac{x^{2b-3}e^{-x}}
{(4x+1)^{1/2}}\ln\left( 1+\kappa x+\frac{\kappa x}{4x+1}\right)
\label{eq:82}
\end{equation} 
The remaining integral (82) over temperatures can be easily evaluated in
different
subregions of the photon momentum $k_{T}$. Below we will be interested
in rather low photon momenta $k_{T}$ (up to $500 Mev$) where transition effect
is expected to be more pronounced. In this momentum region the integral 
(82) depends mainly on final temperature $T_{c}$ (at asymptotically large
momenta $k_{T}$ it depends on initial temperature $T_{0}$).  
The final proper time $\tau_{c}$ in (81) depends on initial conditions
in general. However for two main variants of the initial conditions used
in literature~\cite{NFS}:$\tau_{0}T_{0}=1, T_{0}/T_{c}=3/2 (\tau_{0}=1 fm/c)$
and $\tau_{0}T_{0}=1/3, T_{0}/T_{c}=5/2 (\tau_{0}=0.2 fm/c)$ the proper time  
$\tau_{c}$ changes inessentially, being $3.125 fm/c$ and $3.375 fm/c$
correspondingly (for $b=3$). Below we will use for $\tau_{c}$ an average
value $\tau_{c}=3.25 fm/c$.

Let us note that the photon production rate in (78) can be expressed
through photon occupation number $n(k)$:
\begin{equation}
p_{0}\frac{dR_{\gamma}}{d^{3}p}=\frac{2k_{T}}{(2\pi)^3}\frac{dn(k_{T}\cosh\eta)}
{d\tau}
\label{eq:83}
\end{equation}
(with two polarizations included). In particular,taking (83) into account
one can see
that if the velocities of the volume elements, as well as proper time interval
in the first term in {\sl rhs} of Eq.(78) are small then (78) is reduced to
(56) as it should be. The estimation of (79) shows that
the photon occupation number $n(k)$ in Eq.(83)
is small numerically,
$$
n(k)\ll 1
$$
That means in particular that the transition radiation is dominated not by the
photon amplification but by the ground state rearrangement.

The transition contribution $I^{(1)}_{tr}$ in (78) appears essential
only at small momenta $k_T$. So dealing with single-photon distributions
one can use (76) for evolution parameter $r(k)$. The resulting relative
strength of the transition radiation
\begin{equation}
R_{1}(k_{T})=I^{(1)}_{tr}/I_{QGP}
\label{eq:84}
\end{equation}
appears sizable only in the momentum region $k_{T}\leq (20-30) Mev$ 
independently of the time duration of the transition. It is shown at Fig.1. 

             
Mach better the transition effect is seen in photon correlations
({\sl cf} (64-65))
where it is first order effect with respect to $r(k)$. Let us note that HBT
effect for photons has now more complicated form than that in (64) because
of finite time duration~\cite{APW} of the process of photon emission from QGP  
and it will not be exposed here. We consider only the transition effect (the
third term in (57,64) which gives opposite side correlations) estimating its 
contribution to two-photon correlation function in central rapidity region.
Suggesting fast transition we can evaluate the contribution in the vicinity
of fixed proper time $\tau_c$. So one only has to shift to the rest frame
of each element of the expanding volume and perform $\eta$-integration. Then
the extention of the invariant correlator in (63) to the case of expanding
volume takes the form:
\begin{equation}
2k\langle \tilde b_{\pm}({\vko})\tilde b_{\pm}({\vkt})\rangle 
=G({\bf k}_{1T}+{\bf k}_{2T})I^{(2)}_{tr}
\label{eq:85}
\end{equation}
with
\begin{equation}
I^{(2)}_{tr}=\int d^{2}x_{T}\int d\eta \frac{2\tau_{c}k_{T}\cosh\eta}{(2\pi)^3}
r(k_{T}\cosh\eta)
\label{eq:86}
\end{equation}
where we neglected $n(k)$ in comparison with unity (see above). Therefore
the normalized two-photon correlation function is given by ({\sl cf} (64-65))
\begin{equation}
C({\bf k}_{1T},{\bf k}_{2T})\Bigl|_{y_{1}=y_{2}=0} =1+C_{HBT}
+\frac{1}{2}R^{2}_{2}(k_{T})G^{2}({\bf k}_{1T}+{\bf k}_{2T})
\label{eq:87}
\end{equation}
with
\begin{equation}
R_{2}(k_{T})=\frac{I^{(2)}_{tr}}{I_{QGP}+I^{(1)}_{tr}}
\label{eq:88}
\end{equation}


We calculated the ratio $R_{2}(k_{T})$ for different transition times
$\delta\tau=0, \delta\tau=0.5 fm/c, \delta\tau=1 fm/c$ up to $k_{T}=500 MeV$.
The results are shown at Fig.2.
In the region $k_{T}<100 MeV$ the ratio $R_{2}(k_{T})$ is sizable for all these
$\delta\tau$ 
being equal 4.94 at $k_{T}=0$, reaching maximal value $R\sim 6$ at $k_{T}\sim
20 MeV$ and falling down at $k_{T}=100 MeV$
to $R_{2}=1.78$ for $\delta\tau=0$, $R_{2}=0.95$ for
$\delta\tau=0.5 fm/c$, $R_{2}=0.55$ for $\delta\tau=1 fm/c$ .
At larger transverse momenta the behaviour of the ratio $R$ depends strongly
on transition time $\delta\tau$: in $k_{T}$ interval $(200-500) MeV$
the ratio $R_{2}(k_{T})$ rises for $\delta\tau=0$, it is
approximately constant for $\delta\tau=0.5 fm/c$ and it decreases
for $\delta\tau=1 fm/c$. The asymptotic behaviour of the ratio $R_{2}(k_{T})$
at large photon momenta $k_{T}$ depends on relationship of the temperature
dependent single-particle production rate (decreasing like $\exp(-\beta k)$
and the transition time dependent evolution parameter (decreasing like
$\exp(-\pi k\delta\tau/2)$ in our reference model). In any case, one can hope
to see the effect of the transition in the region $k_{T}\le 100 Mev$
where the peak of the ratio $R_{2}(k_{T})$ is always present.

\section{Conclusion}
Estimation of photon emission accompaning transition between quark-gluon and
hadron states of matter in heavy ion collisions shows that opposite side
photon correlations can serve as a sign of the transition if transition
time is not large.
  
\section{Acknowledgments}
The work was supported by the Russian Fund for Basic Research (grant 00-02-101).

\newpage
{\large \bf Figure captions}

\vspace {1 cm}

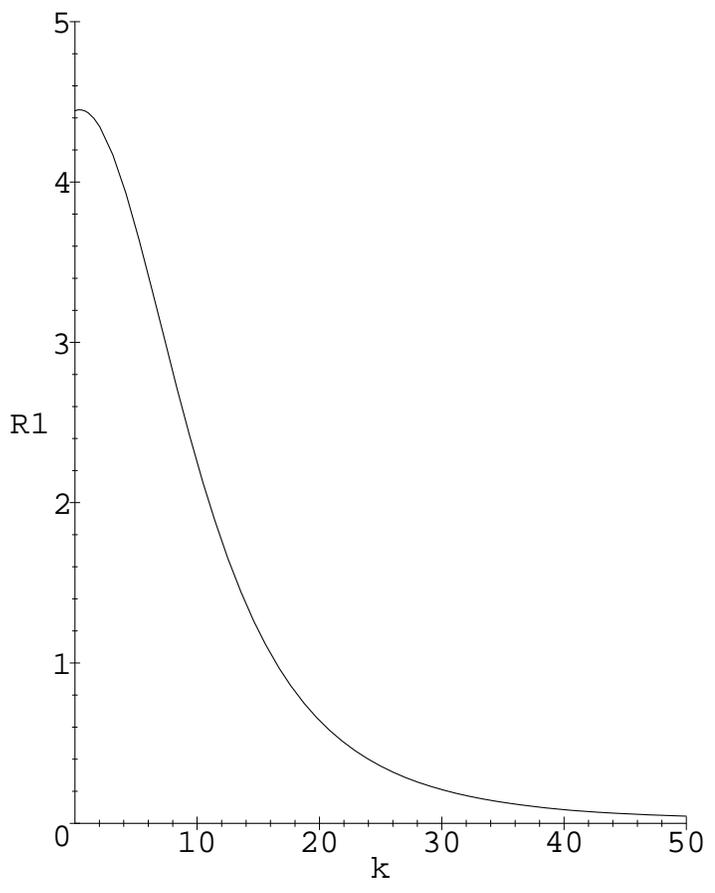
{\bf Fig.1} The relative strength of the transition radiation for transition
from QGP to hadrons.

\vspace {0.5 cm}

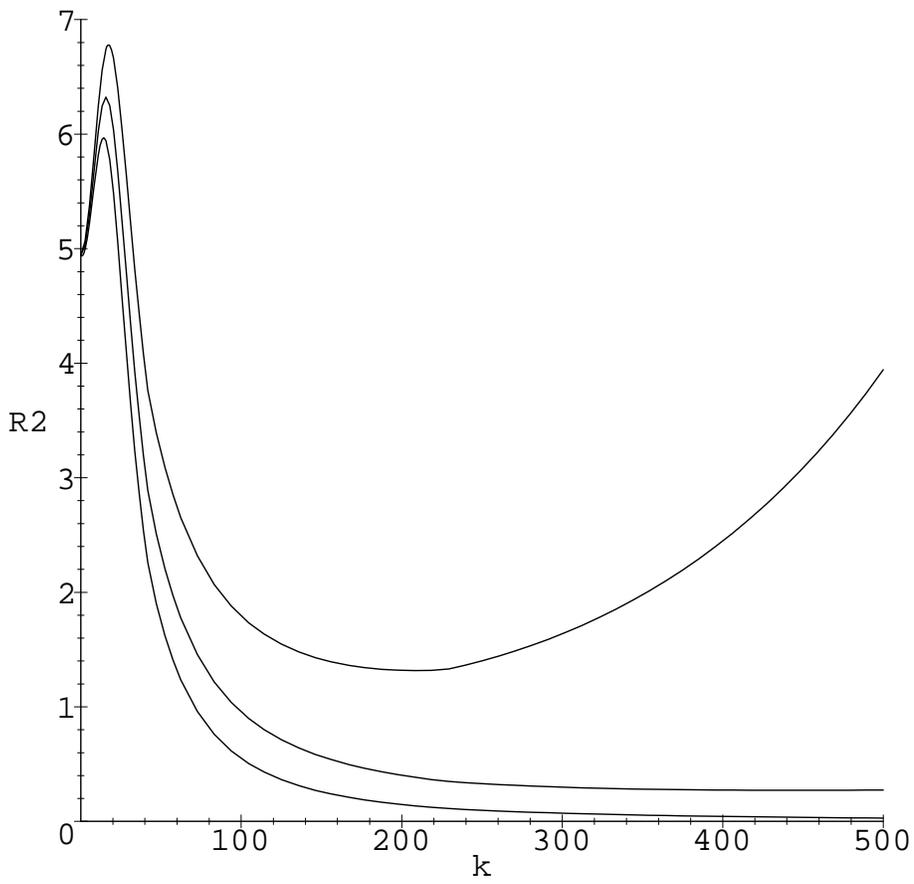
{\bf Fig.2} The relative strength of the opposite side two-photon correlations
for transition time $\delta\tau = 0, \delta\tau = 0.5fm/c, \delta\tau = 1fm/c $
(from top to bottom), see Eqs.87-88.

\end{document}